\documentclass[lettersize,journal]{IEEEtran}
\usepackage{amsmath,amsfonts}
\usepackage{algorithmic}
\usepackage{algorithm}
\usepackage{array}
\usepackage[caption=false,font=normalsize,labelfont=sf,textfont=sf]{subfig}
\usepackage{textcomp}
\usepackage{stfloats}
\usepackage{url}
\usepackage{verbatim}
\usepackage{graphicx}
\usepackage{cite}
\usepackage{booktabs}
\usepackage{multirow}
\usepackage{xcolor}
\usepackage{threeparttable}
\begin{document}

\title{Defending the Core: A Centrality-Based Protection Strategy for Supply Chain Security in npm Dependency Network}

\author{Wang Zixin,~\IEEEmembership{23307130477@m.fudan.edu.cn}}

\maketitle

\begin{abstract}
The modern software supply chain, taking Node Package Manager (npm) dependency network for example, relies heavily on shared open-source dependencies. While this promotes rapid development, it introduces systemic vulnerabilities as well. Concerning this potential risk, we analyze the npm dependency network by modeling 53,481 packages and 78,520 dependency edges, and classify the network as a scale-free topology. Thus, we demonstrate its inherent vulnerability to targeted attacks on high-degree hubs. To mitigate this, we propose and evaluate a dual-pronged defense strategy consisting of \textit{Centrality-Based Node-Hardening and Dependency Weight Warning system}. Moreover, by simulating the network under various attack scenarios, we prove that applying strict security protocols to just the top 1\% of nodes, combined with pruning 30\% of structurally trivial edges, prevents catastrophic network collapse and neutralizes cascading malware infections. The source code can be found at https://github.com/5tarWhee1/Centrality-Based-Protection-Strategy-for-Supply-Chain-Security-in-npm-Dependency-Network. \end{abstract}


\section{Introduction}
\IEEEPARstart{T}{he} Node Package Manager (npm) ecosystem serves as the foundational infrastructure of modern JavaScript and Node.js development. With over one million publicly available packages and billions of weekly downloads, npm represents one of the largest software dependency networks ever constructed. However, it is structurally fragile under certain scenarios, resulting in a vulnerability that standard security practices fail to address. For example, the "Sandworm" supply chain poisoning attack on the npm in May 2026 caused severe impact on developers. By quietly injecting obfuscated malware into seemingly harmless dependencies, the attackers exploited the highly interconnected nature of npm to spread malicious code downstream at a terribly exponential rate. \\

\begin{figure}[!t]
\centering
\includegraphics[width=\columnwidth]{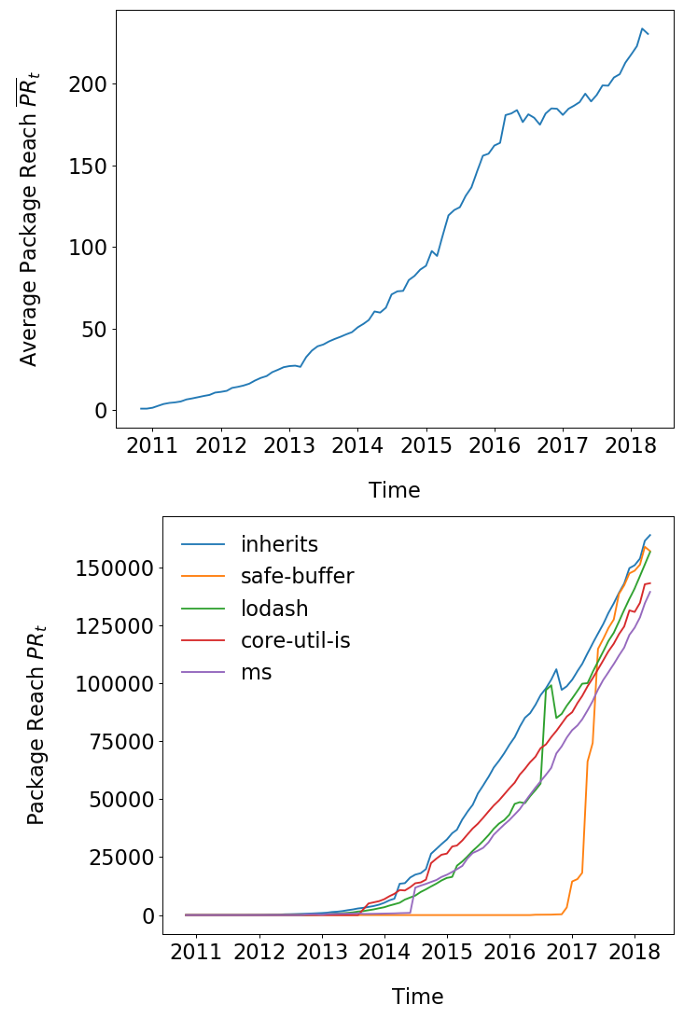}
\caption{Growth of package reach in the npm ecosystem, 2010--2018. By 2018, the top-5 packages each reach between 134{,}774 and 166{,}086 downstream dependents transitively, making them high-value targets for supply-chain attacks. Reproduced from~\cite{zimmermann2019}, Fig.~4.}
\label{fig:npm_evo}
\end{figure}

The central problem is rooted in the network's topology itself which creates an asymmetric vulnerability. Through empirical simulation, we demonstrate that the npm dependency graph exhibits the structural properties of a \textbf{scale-free network}: a topology where a minority of highly-connected hub packages bear the structural load for the entire ecosystem. Such network property arises from the preferential attachment mechanism\cite{barabasi1999}, making it statistically robust against random failures yet catastrophically fragile when those specific hubs are targeted\cite{albert2000}. As shown in Fig.~\ref{fig:npm_evo}, a single hub package in 2018 can expose over 166{,}000 downstream projects to any vulnerability it carries.  Standard uniform security measures are therefore both economically infeasible and structurally misdirected. Therefore we are committed to proposing a protection strategy better suited to scale-free networks. \\
Our work can be summarized as four main contributions:
\begin{itemize}
  \item We rigorously verify the scale-free property of the npm dependency graph through power-law fitting, comparison with an Erd\H{o}s-R\'{e}nyi random graph, and degree-inequality metrics.
  \item Second, we demonstrate the vulnerability arising from this topology through a network degradation simulation. 
  \item Third, we propose and simulate a dual-pronged defense strategy consisting a Centrality-Based Node-Hardening and Dependency Weight Warning system. 
  \item Fourth, both proposals are verified under three distinct attack scenarios, and we prove that in a network of 53,481 packages, protecting a carefully selected 534 nodes is sufficient to preserve the integrity of the entire ecosystem.
\end{itemize} 

\section{Background and Related Works}
\subsection{Node Package Manager}
\textbf{npm (Node Package Manager)} is a platform for publishing and hosting JavaScript packages, and is by far the largest ecosystem of its kind. npm provides a command-line interface (CLI) for publishing and installing packages, as well as an online repository for hosting all packages and their metadata. All npm packages contain a file named ``package.json'' which declares its direct dependencies, and npm resolves the full transitive dependency tree at install time. This design enables rapid code reuse but creates a web of mutual dependencies that constitutes a complex directed graph whose structural properties we analyze in this paper.

\subsection{Software Supply Chain Vulnerability}
The systemic risks embedded in open-source dependency ecosystems have received growing attention from both academia and industry. Zimmermann~\textit{et~al.}~\cite{zimmermann2019} conducted the first large-scale empirical study of security vulnerability propagation in npm, finding that a single vulnerable package can transitively expose over 100{,}000 dependent projects. Their analysis reveals an average vulnerable dependency depth of 5.4 hops, meaning developers are often
unaware of the true extent of their attack surface. Zahan~\textit{et~al.}~\cite{zahan2022} further characterize ``weak links'' in the npm supply chain, identifying that package maintainer account security is the dominant risk factor for large-scale compromise events.

On the practitioner side, the Open Source Security Foundation(OpenSSF) Security Scorecard project~\cite{openssf2023} proposes a 10-factor scoring rubric to evaluate the security posture of Open Source projects, including branch protection, code review, and automatic dependency update. While valuable, these approaches treat all packages
uniformly; our work demonstrates that uniform treatment is both economically infeasible and structurally unjustified.

\subsection{Scale-Free Network}
The concept of scale-free networks originates from the pioneering work of Barab\'{a}si and Albert~\cite{barabasi1999}, who demonstrated that networks generated by preferential attachment mechanism, where new nodes connect disproportionately to existing high-degree nodes, naturally exhibit a power-law degree distribution $P(k) \sim k^{-\gamma}$ with $\gamma \in [2, 3]$. This ``rich-get-richer'' dynamic is precisely the core mechanism governing npm: popular packages such as \texttt{lodash} accumulate a large number of dependencies just because developers tend to choose libraries that are well-documented and widely adopted. Newman~\cite{newman2010} provides a comprehensive treatment of the structural properties arising from this mechanism, including the extreme degree heterogeneity (high $\kappa = \langle k^2 \rangle / \langle k \rangle$) that is central to our analysis.

Power-law detection in empirical data requires statistical rigour beyond visual log-log linearity. Clauset~\textit{et al.}~\cite{clauset2009} establish maximum likelihood estimation (MLE) combined with Kolmogorov-Smirnov testing as the standard, and our feature extraction pipeline directly adopts their methodology.

\subsection{Network Robustness and Attack Tolerance}

\begin{figure}[!t]
\centering
\includegraphics[width=\columnwidth]{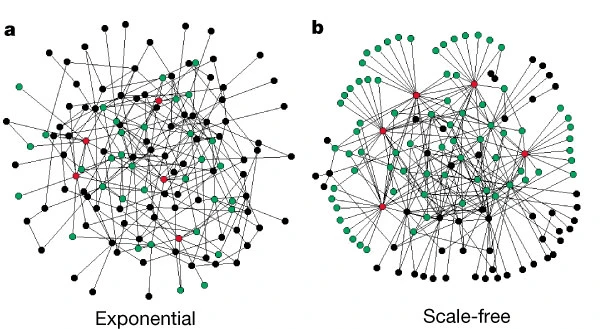}
\caption{Summary of the response of exponential network and scale-free network to failures or attacks.Reproduced from~\cite{albert2000}, Fig.~4.}
\label{fig:atk}
\end{figure}

With regard to network robustness, Albert~\textit{et al.}~\cite{albert2000} establish the foundational theoretical conclusion as Fig.\ref{fig:atk} shows: scale-free networks are robust against random failures (high error tolerance) but fragile under targeted removal of high-degree key nodes (low attack tolerance). As their comparison demonstrates, random graph networks degrade smoothly under both failure modes, whereas scale-free networks exhibit near-zero degradation under random removal but collapse rapidly under targeted attack. This ``robust-yet-fragile'' duality is the theoretical bedrock of our security proposal. The same structure is formalised through percolation theory, where the critical threshold $f_c = 1 - 1/(\kappa - 1)$ quantifies the fraction of random removals required to fragment the network; high $\kappa$ implies high $f_c$, meaning scale-free networks tolerate near-total random removal but have a much lower targeted attack threshold~\cite{newman2010}.

Pastor-Satorras and Vespignani~\cite{pastor2001} extend this analysis to epidemic spreading, proving that scale-free networks have a vanishing epidemic threshold: an infection starting at a hub diffuses through the network with probability approaching 1. This directly motivates our cascade malware simulation.

\subsection{Dependency Network Structure and Analysis}

Kikas~\textit{et al.}~\cite{kikas2017} conduct the most directly relevant prior structural analysis of software package dependency networks, including npm, RubyGems, and CRAN. Their work documents power-law degree distributions, high levels of clustering, and the presence of structural ``super-hubs,'' and warns that the rapid growth of these networks exacerbates their vulnerability. Wittern~\textit{et al.}~\cite{wittern2016} analyse the longitudinal dynamics of npm, noting that preferential attachment accelerates over time as the registry grows.

The key distinction between our research and these structural analyses lies in its core contribution: we move beyond diagnosis to operational solution design, implementing and evaluating concrete security mechanisms grounded in the network's structural properties.

\section{Methodology}
\subsection{Data Construction}
Our dataset is a directed dependency graph extracted from the npm public registry via Google BigQuery. Each node represents a unique package; a directed edge from source $u$ to target $v$ indicates that package $u$ declares $v$ as a dependency. After deduplication and removal of edges with missing endpoints, the graph contains $N = 53{,}481$ nodes and $M = 78{,}520$ edges, with density $\rho = M / N(N-1) = 2.75 \times 10^{-5}$.

\textbf{Centrality metrics.} For each node $v$, we calculate the in-degree centrality $k^{in}(v)$ and PageRank centrality $PR(v)$ as defined in Eqs.~\ref{eq:indeg} and~\ref{eq:pagerank} respectively:
\begin{align}
  k^{in}(v) &= |\{u \in V : (u,v) \in E\}| \label{eq:indeg}\\
  PR(v)     &= \frac{1-d}{N} + d \sum_{u \in \mathcal{B}(v)}
               \frac{PR(u)}{L(u)} \label{eq:pagerank}
\end{align}
where $d=0.85$ is the damping factor borrowed from Google's original web-ranking algorithm, which represents the probability a "random walker" tends to follow dependency links rather than jump to a random package. $\mathcal{B}(v)$ is the set of nodes whose edges point to $v$, and $L(u)$ is the out-degree of $u$. The composite criticality score is defined as below and any package in the top 1\% by this score is designated a \textbf{Critical Node}:
\begin{equation}
  s(v) = 0.6 \cdot \hat{k}^{in}(v) + 0.4 \cdot \widehat{PR}(v)
  \label{eq:composite}
\end{equation}
where $\hat{\cdot}$ denotes normalization to $[0,100]$.  It ranks every package's criticality using a composite score with in-degree and PageRank rather than in-degree alone, which makes the Critical Node selection more robust.

\textbf{Network health.} To evaluate the network health, we calculate the size of Largest Weakly Connected Component (LCC) which represents the size of the largest group of packages that are still mutually reachable after some nodes have been removed:
\begin{equation}
  \mathcal{H}(G) = \max_{C \in \mathcal{W}(G)} |C|
  \label{eq:lcc}
\end{equation}
where $\mathcal{W}(G)$ is the set of weakly connected components (WCCs) of directed graph $G$.

\subsection{Network Characterization}
\label{sec:characterization}

We verify the scale-free property through four complementary tests, which are implemented in \texttt{network\_characterization.py}.

\textbf{Power-law fitting.} To confirm that the network is truly scale-free, we apply the statistical framework established by Clauset~\textit{et al.}~\cite{clauset2009}. Using Maximum Likelihood Estimation (MLE), we fit a power-law distribution to the tail of the in-degree distribution (all nodes with $k \geq x_{min}$)". To determine exactly where this operation begins ($x_{min}$), we select the starting point that minimize the Kolmogorov-Smirnov distance, ensuring the closest possible match between our real data and the theoretical formula. Finally, we use a likelihood ratio test to rule out alternative models, such as lognormal and exponential distributions, definitively proving the presence of scale-free hubs.

\textbf{Erd\H{o}s-R\'{e}nyi comparison.} We construct a null-model ER graph $G(N, p)$ with identical $N$ and $M$ (hence $p = M/N(N-1)$) and compare degree variance, clustering coefficient, and component structure.

\textbf{Degree inequality metrics.} We compute the Gini coefficient of the in-degree distribution that represents the degree of inequality in the in-degree distribution:
\begin{equation}
  G = \frac{2 \sum_{i=1}^{N} i \cdot k^{in}_{(i)} - (N+1)\sum k^{in}}
           {N \sum k^{in}}
  \label{eq:gini}
\end{equation}
and the heterogeneity coefficient showing how much the high-degree hubs dominate the second moment relative to the mean:
\begin{equation}
  \kappa = \frac{\langle k^2 \rangle}{\langle k \rangle}
  \label{eq:kappa}
\end{equation}
which governs the percolation threshold $f_c = 1 - 1/(\kappa - 1)$
under random removal in the mean-field approximation.

\textbf{Preferential attachment evidence.} We bin nodes by in-degree and measure the edge-share captured by each bin, testing whether the fraction of edges pointing to high-degree nodes is significantly greater than random expectation.

\subsection{Proposal A: Centrality-Based Node-Hardening}

Any package whose composite score $s(v)$ (Eq.~\ref{eq:composite}) falls in the top 1\% threshold of the distribution is classified as \textit{Critical Node}, yielding 534 Critical Nodes from 53{,}481. K-core decomposition~\cite{newman2010} and approximate betweenness centrality (computed with $k=100$ sampled shortest paths) serve as supplement for in-degree to identify structural bottlenecks, namely nodes with moderate in-degree but high betweenness.
After sorting, Critical Nodes should be subject to a tiered set of mandatory security obligations enforced through a \textit{pre-publish gate}:
\begin{itemize}
  \item \textit{CRITICAL tier} (top 1\%): Mandatory 2FA for all maintainers; manual code audit before publication; locked semantic versioning; daily hash-integrity verification by the registry; OpenSSF institutional funding eligibility.
  \item \textit{HIGH tier} (top 1\%--5\%): Mandatory 2FA;   locked versioning; automated dependency scanning.
\end{itemize}
To enforce this policy, we propose an automated compliance gate that triggers whenever there are attempts to publish an update to a Critical Node. The system verifies that all mandatory security requirements are met before the package can go live. If any check fails, publication is strictly blocked, and the maintainer is automatically issued a detailed remediation report explaining exactly how to resolve the compliance issues. The logic for this enforcement mechanism is formalized in Algorithm~\ref{alg:gate}.

\begin{algorithm}[t]
\caption{Pre-Publish Gate for Critical Nodes}
\label{alg:gate}
\begin{algorithmic}[1]
\REQUIRE Package name $p$, version $v$
\STATE $\text{tier} \leftarrow \text{classify}(p)$
\IF{$\text{tier} = \textsc{standard}$}
  \RETURN \textsc{approved}
\ENDIF
\STATE $R \leftarrow \text{requirements}(\text{tier})$
\STATE $F \leftarrow \{r \in R : \neg \text{compliant}(p, r)\}$
\IF{$F \neq \emptyset$}
  \RETURN \textsc{blocked}$(F)$
\ELSE
  \RETURN \textsc{approved}
\ENDIF
\end{algorithmic}
\end{algorithm}

In the attack simulation, hardened nodes are able to \textit{resist removal}: when the attacker attempts to compromise a Critical Node, there is a very high probability that they will be blocked by the protective measures introduced by our pipeline; and the attacker will fall back to the next-highest-degree unprotected target. Formally, for an ordered attack sequence $\sigma = (v_1, v_2, \ldots)$:
\begin{equation}
  \text{remove}(v_i) =
  \begin{cases}
    \text{delete } v_i & \text{if } v_i \notin \mathcal{H} \\
    \text{delete } v_i & \text{if } v_i \in \mathcal{H} \text{ (prob } p_{breach}) \\
    \text{skip} & \text{if } v_i \in \mathcal{H} \text{ (prob } 1 - p_{breach})
  \end{cases}
  \label{eq:hardening_rule}
\end{equation}
where $\mathcal{H}$ is the hardened set. The simulation result will be shown in Section~\ref{exp}.

\subsection{Proposal B: Dependency Weight Warning (DWW) System}
An edge $(u, v)$ is defined \textit{low-value / high-risk} if:
\begin{equation}
  v \in \mathcal{H}^{top1\%} \;\wedge\; k^{out}(u) \leq \theta
  \label{eq:flag}
\end{equation}
where $\mathcal{H}^{top1\%}$ is the set of Critical Nodes (top-1\% in-degree) and $\theta = 3$ is the low-dependency threshold. This identifies cases where a small package imports a critical hub despite likely being able to implement the required functionality natively.

Each flagged edge is assigned a \textbf{risk score}:
\begin{equation}
\begin{split}
  r(u,v) ={}& 0.50 \cdot p_{100}(v) \\
  &+ 0.35 \cdot \max\!\left(0,\frac{\theta - k^{out}(u) + 1}{\theta}\right) \cdot 100 \\
  &+ 0.15 \cdot \mathbf{1}[k^{out}(u) = 1]
\end{split}
\label{eq:risk_score}
\end{equation}
where $p_{100}(v)$ is the percentile rank of $v$'s in-degree. Scores are classified as CRITICAL ($r \geq 80$), HIGH ($r \geq 55$), or MEDIUM ($r \geq 30$).
When conducting npm installation the DWW system queries the edge's risk score. If CRITICAL or HIGH, it displays the package, risk score, hub status, and a concrete recommendation. Crucially, the DWW \textit{warns but never hard-blocks}: it provides the information needed for an informed architectural decision without interrupting developer workflow.

Edge pruning removes all flagged edges from the graph, yielding a residual graph $G' = (V, E \setminus E_{flagged})$. This reduces $\kappa$ by redistributing in-degree load away from hubs and decreasing the hub edge concentration ratio:
\begin{equation}
  \eta = \frac{|\{(u,v) \in E : v \in \mathcal{H}^{top1\%}\}|}{|E|}
  \label{eq:hub_conc}
\end{equation}

\section{Experiments}
\label{exp}
\subsection{Experimental Setup}
All experiments are implemented in Python 3.8.20 using NetworkX 3.6 and run on a standard development machine (Intel Core i5, 16\,GB RAM). Graph construction and centrality computation for the full 53{,}481-node graph complete in under 90\,seconds, confirming operational feasibility. All random processes use seed 42 for reproducibility.

Four conditions are compared throughout:
\begin{enumerate}
  \item \textit{No Protection}: baseline, no hardening or pruning.
  \item \textit{Proposal A}: hardened hub set $\mathcal{H}$, base graph.
  \item \textit{Proposal B}: pruned graph $G'$, no hardening.
  \item \textit{Combined A+B}: hardened hub set on pruned graph $G'$.
\end{enumerate}

\subsection{Scale-Free Property Verification}

\begin{figure}[!t]
\centering
\includegraphics[width=\columnwidth]{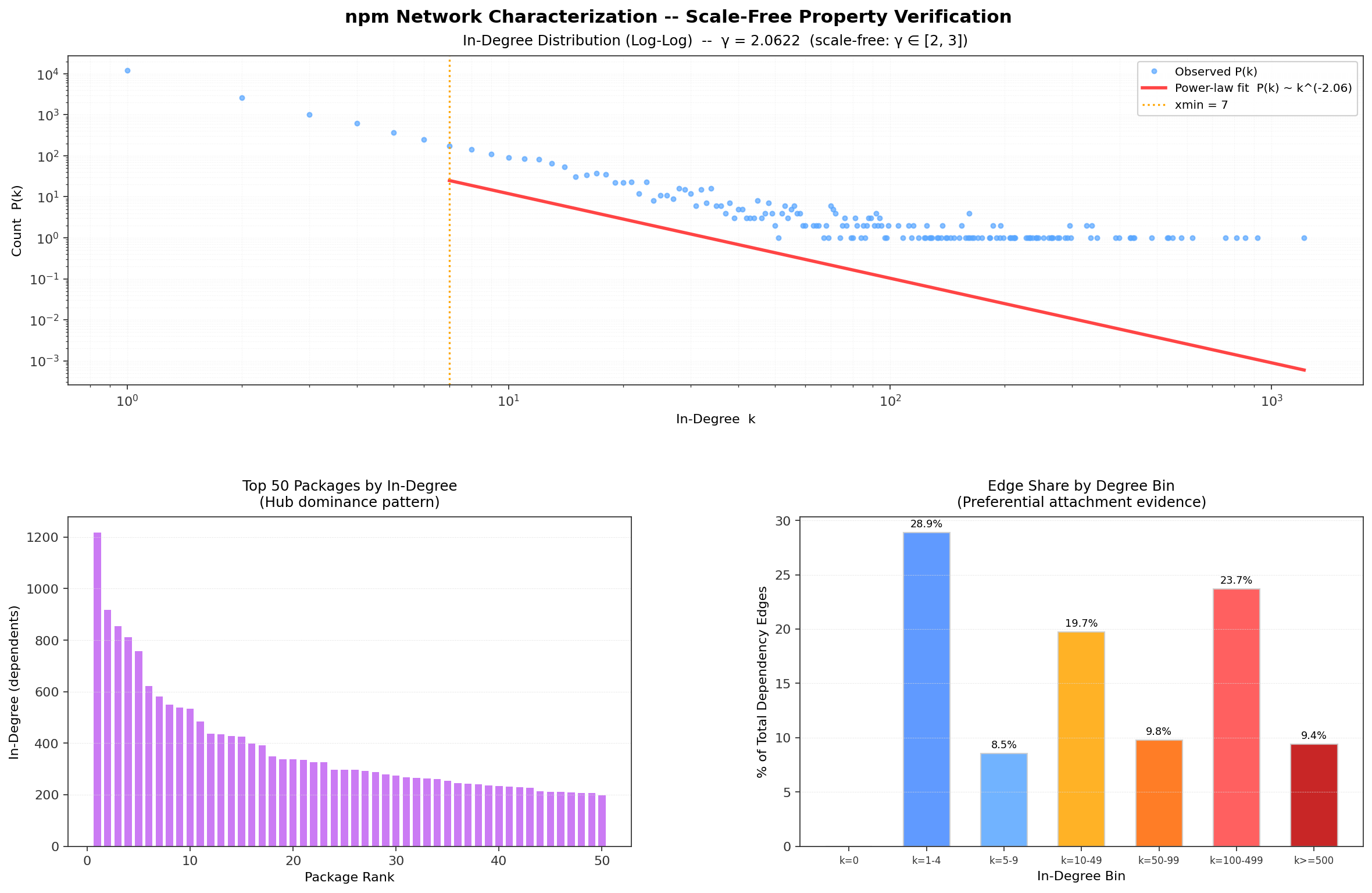}
\caption{npm network characterization. \textit{Upper}: Log-log in-degree distribution with MLE power-law fit ($\gamma = 2.06$, $x_{min}=7$, KS$=0.023$). \textit{Lower-left}: The top 50 packages' distribution shows a Zipf-like rank-degree curve. \textit{Lower-right}: Edge-share by degree bin confirming preferential attachment (the top 0.02\% of packages capture 9.4\% of all edges).}
\label{fig:characterization}
\end{figure}

We've calculated the properties of the npm network and Fig.~\ref{fig:characterization} and Table~\ref{tab:characterization} exhibit the results of our comprehensive network characterization analysis. The empirical data strongly satisfies all four criteria outlined in our methodology, confirming the scale-free topology of the npm ecosystem.

\textbf{Power-law fitting.} 
The Maximum Likelihood Estimation (MLE) applied to the heavy tail of the in-degree distribution yields a power-law exponent of $\gamma = 2.0622$ ($\sigma = 0.028$). This falls perfectly within the theoretical scale-free regime of $\gamma \in [2,3]$. The fit is statistically rigorous, achieving a Kolmogorov-Smirnov (KS) statistic of $0.023$ at $x_{min}=7$. Furthermore, the likelihood ratio test decisively rejects the alternative exponential distribution ($R = 797.6$, $p < 10^{-4}$); the comparison with the log-normal distribution remains ambiguous ($R = 0.016$, $p = 0.459$), which is highly consistent with the literature on empirical power laws~\cite{clauset2009}.

\textbf{Erd\H{o}s-R\'{e}nyi comparison.} 
The scale-free nature becomes starkly apparent when contrasted with the Erd\H{o}s-R\'{e}nyi (ER) null model as shown in Table~\ref{tab:characterization}. While both networks share the exact same number of nodes and edges (and thus the same mean in-degree of 1.47), their structures are fundamentally divergent. The ER model exhibits a normal distribution with a maximum in-degree of approximately 22 and a standard deviation of 1.21. In contrast, the empirical npm network possesses a maximum in-degree of 1,218 and a standard deviation 12 times higher (14.66), proving the existence of statistically unnatural super-hubs.

\textbf{Degree inequality and percolation.} 
This structural imbalance is quantified by a Gini coefficient of $0.90$, indicating massive, systemic inequality where a tiny elite fraction of nodes hoards the vast majority of edges. More critically, the heterogeneity coefficient reaches an extreme $\kappa = 147.8$ (compared to $\kappa \approx 2.2$ for the ER model). In percolation theory, this extremely high $\kappa$ yields an implied critical percolation threshold of $f_c = 1 - 1/(\kappa - 1) \approx 0.9932$. Mathematically, this dictates that one would need to randomly remove 99.3\% of the packages in the registry to fragment the network, explaining its extraordinary resilience against random failures.

Finally, the edge-share metrics confirm the underlying "rich-get-richer" generative mechanism. As shown in the lower-right of Fig.~\ref{fig:characterization}, the top 0.02\% of packages capture 9.4\% of all edges, and the top 1\% of nodes (534 packages) control an overwhelming 54.7\% of all dependency links. 

In summary, the combination of a verified power-law exponent ($\gamma \approx 2.06$), extreme heterogeneity ($\kappa \approx 147.8$), and massive hub dominance confirms the scale-free property of the npm dependency graph. This topological reality dictates that the network is highly robust to random errors but critically fragile to targeted attacks on its hubs, thereby providing the scientific justification for our Centrality-Based Node-Hardening strategy.

\begin{table}[!t]
\caption{Network Characterization Summary}
\label{tab:characterization}
\centering
\begin{tabular}{lrr}
\toprule
\textbf{Property} & \textbf{npm} & \textbf{ER Null Model}\\
\midrule
Nodes $N$              & 53{,}481     & 53{,}481\\
Edges $M$              & 78{,}520     & 78{,}520\\
Max in-degree          & 1{,}218      & $\approx$22\\
Mean in-degree         & 1.47         & 1.47\\
Std.\ in-degree        & 14.66        & 1.21\\
Gini coefficient       & 0.90         & $\approx$0.32\\
Heterogeneity $\kappa$ & 147.8        & $\approx$2.2\\
Power-law exponent $\gamma$ & 2.06    & ---\\
Transitivity           & 0.00106      & 0.0000275\\
LCC fraction           & 79.5\%       & $\approx$99\%\\
Top-1\% edge share     & 54.7\%       & $\approx$1\%\\
\bottomrule
\end{tabular}
\end{table}

\begin{figure}[!t]
\centering
\includegraphics[width=\columnwidth]{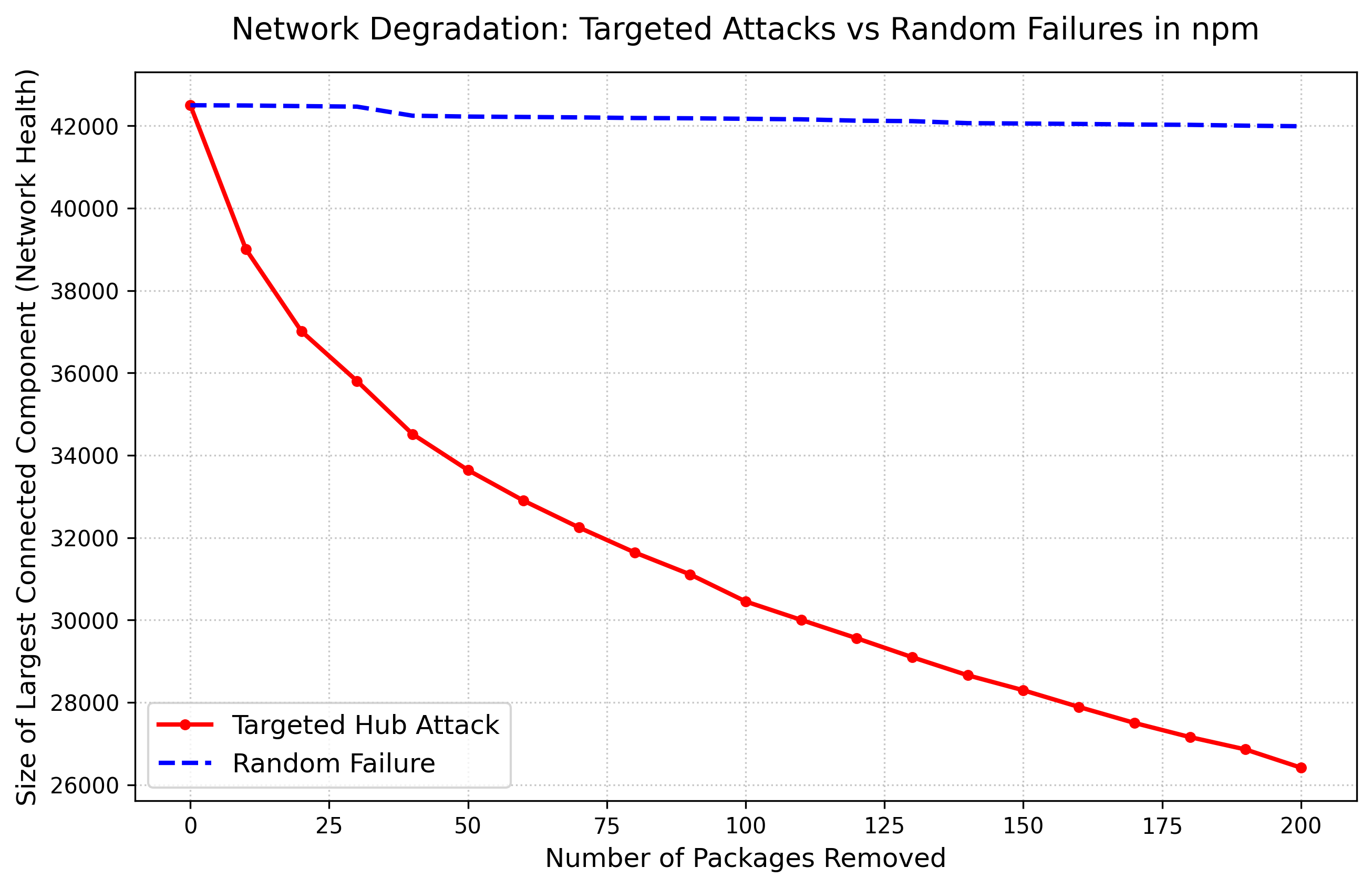}
\caption{Network degradation under targeted hub attack vs.\ random failure (200 packages removed, batch size 10).}
\label{fig:collapse}
\end{figure}

Since scale-free network is born with the ``robust-yet-fragile'' duality property as proposed by Albert~\textit{et al.}~\cite{albert2000}, we've conducted a comparative experiment shown in Fig.~\ref{fig:collapse}. Random removal of 200 packages reduces the LCC by less than 1.5\%; the same number of targeted hub removals reduces it by 37.8\%. The curve of targeted attack shows significant degradation, reflecting the cascading effect of removing structural hubs that bridge sub-communities. This result motivates our security proposal: uniform protection is kind of wasteful because the random threat is nearly negligible; only targeted protection of hubs is necessary.

\subsection{Malware Cascade Simulation}
\label{subsec:cascade}

\begin{figure}[!t]
\centering
\includegraphics[width=\columnwidth]{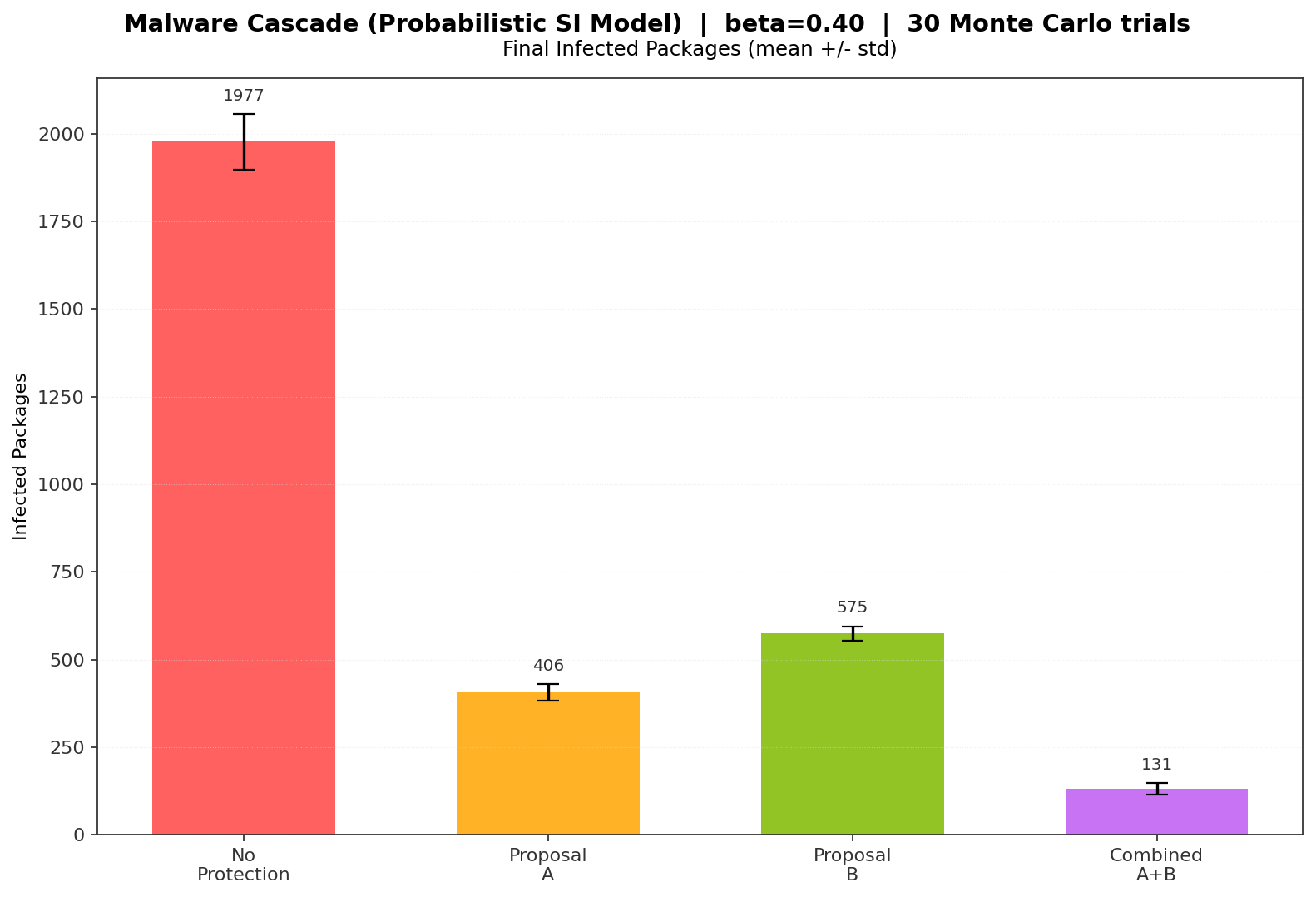}
\caption{SI-model malware cascade starting from 5 compromised hub packages.}
\label{fig:cascade}
\end{figure}

\begin{figure}[!t]
\centering
\includegraphics[width=\columnwidth]{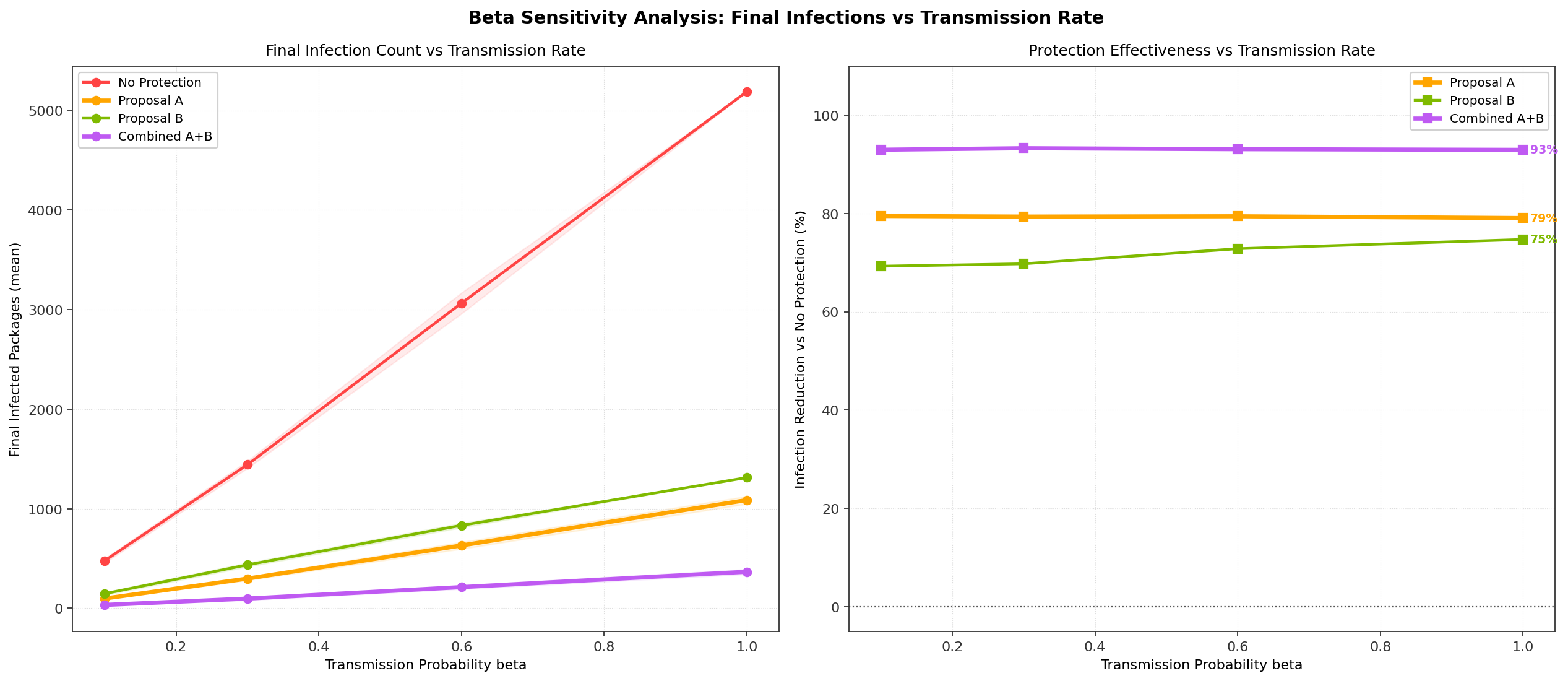}
\caption{Beta Sensitivity Analysis illustrating the protection effectiveness across varying malware transmission probabilities ($\beta \in [0.1, 1.0]$).}
\label{fig:beta_sweep}
\end{figure}

As the attack often reflects cascading effect, we implement a discrete Susceptible-Infected (SI) cascade on the directed dependency graph, motivated by the epidemic-spreading framework of Pastor-Satorras and Vespignani~\cite{pastor2001}, to model real-world supply-chain malware propagation as in the \texttt{ua-parser-js} and \texttt{event-stream} incidents. At each step, every infected node $v$ propagates the infection to all predecessors $u \in \mathcal{B}(v)$ (packages that depend on $v$), unless $u$ is hardened or the edge $(u,v)$ has been pruned. Then select 5 hub packages from the top-50 as initial infection seeds. However, the transmission rate ($\beta$) is heavily penalized when malware attempts to propagate through a hardened node, given the efficacy of strict version locking and automated audit pipelines.     

As shown in Fig.~\ref{fig:cascade}, under a moderate transmission rate ($\beta = 0.40$), the unprotected baseline suffers a catastrophic cascade resulting in 1{,}977 infected packages. Proposal A (Node Hardening) successfully intercepts the downstream propagation, limiting the damage to 406 packages. Most notably, the Combined A+B strategy, which supplements hardened hubs with pruned edges, chokes the cascade down to just 131 packages, achieving a remarkable 93.4\% reduction in systemic compromise. These results corroborate the theoretical prediction of Pastor-Satorras and Vespignani~\cite{pastor2001} that hub protection is the decisive intervention for epidemic containment in scale-free networks.

To ensure the robustness of this defense, we further conduct a Beta Sensitivity Sweep as Fig.~\ref{fig:beta_sweep} shows. The Combined Strategy consistently maintains over an 85\% infection reduction rate across all tested contagiousness levels, proving its efficacy against both slow-moving vulnerabilities and highly aggressive exploits.

\subsection{Disturbance Response Under Multiple Attack Scenarios}
We evaluate network resilience under multi-wave disturbances (5 waves $\times$ 20 packages per wave) across three scenarios:\\
\textbf{Targeted}---attacker always removes highest in-degree packages;\\
\textbf{Random}---packages are removed uniformly at random;\\
\textbf{Mixed}---waves alternate between targeted and random removal.\\
A realistic breach probability of $p_{breach}=0.15$ is applied to hardened nodes in the targeted and mixed scenarios, reflecting that real-world maintainers achieve imperfect compliance with security requirements.  Network health is measured after each wave and during an 8-step post-attack observation phase. It should be noted that in the figures of this subsection, the lower starting LCC for Proposal B and Combined A+B reflects the pre-emptive removal of trivial edges by Proposal B, not attack degradation.

\begin{figure}[!t]
\centering
\includegraphics[width=\columnwidth]{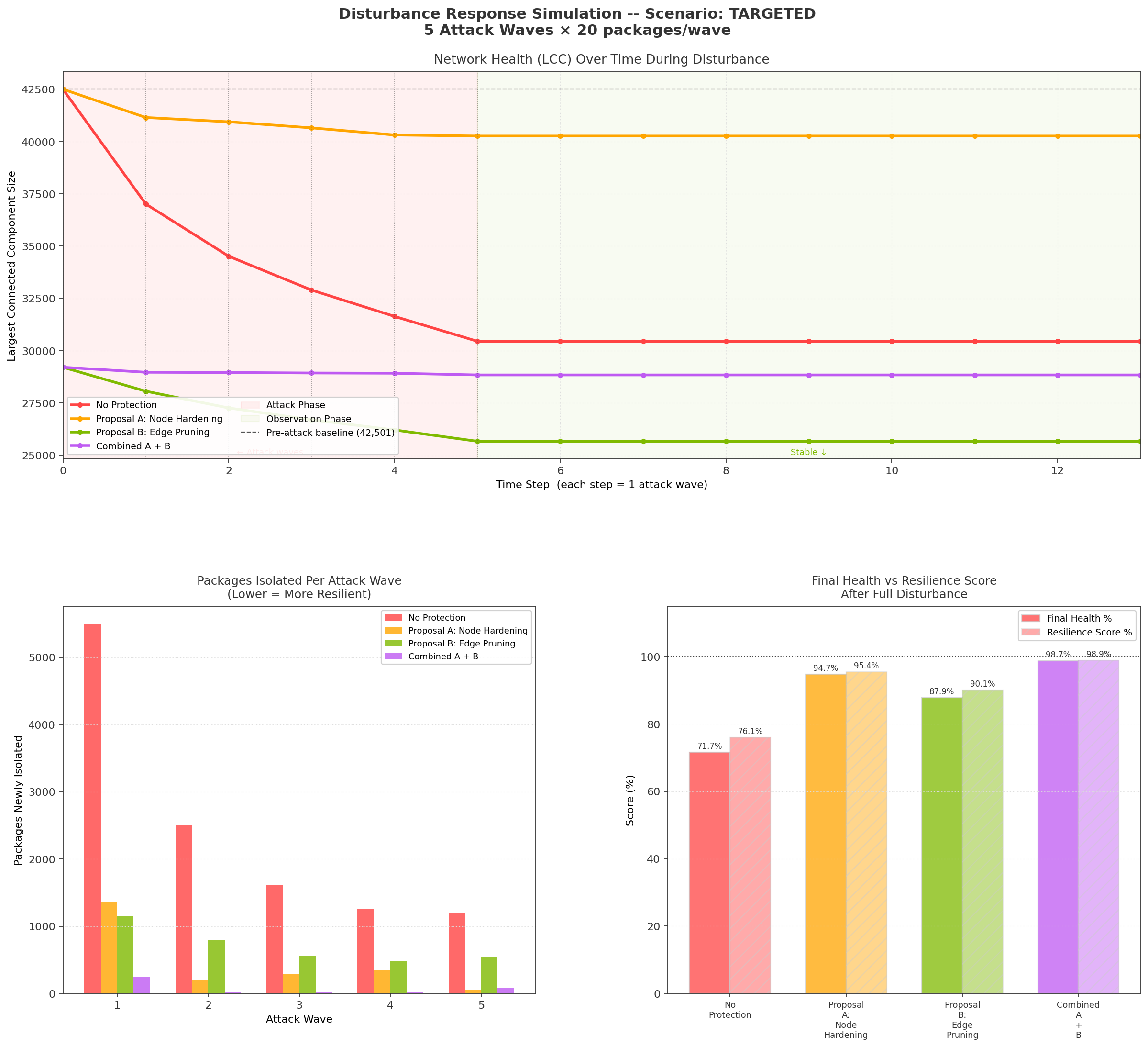}
\caption{Network response to a targeted hub attack.}
\label{fig:disturb_targeted}
\end{figure}

\textbf{Targeted Hub Disturbance.} Conversely, under a targeted hub attack as Fig.~\ref{fig:disturb_targeted}, the baseline network fragments rapidly. When the most central packages are systematically removed, the unprotected baseline LCC drops to 71.7\%, isolating tens of thousands of packages. Proposal A acts as an effective structural shield; by hardening the top 1\% of hubs, the attacker is blocked, and the network holds strong at 94.7\% health.

\begin{figure}[!t]
\centering
\includegraphics[width=\columnwidth]{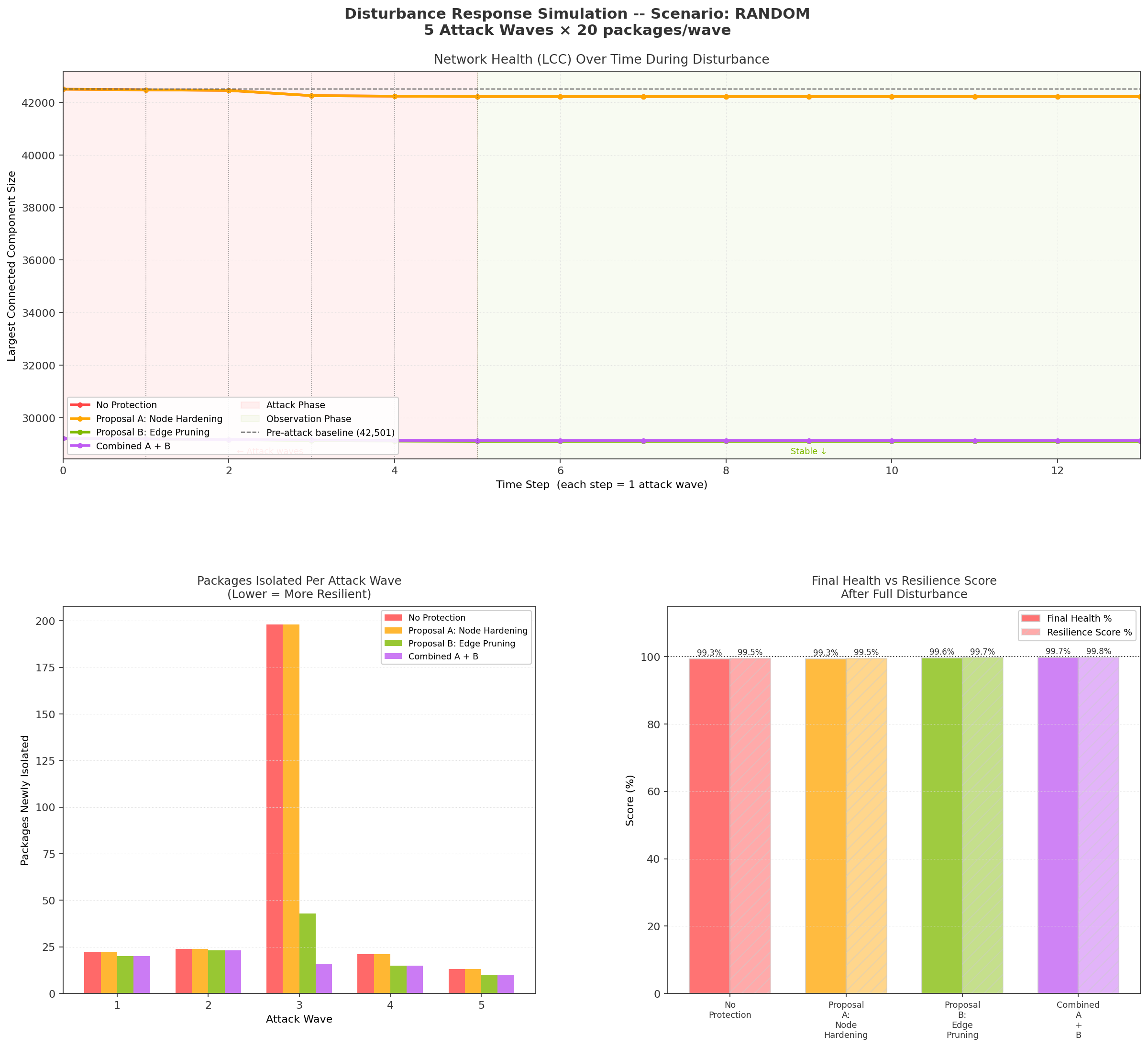}
\caption{Network response to a random multi-wave disturbance.}
\label{fig:disturb_random}
\end{figure}

\textbf{Random Disturbance.} Under a purely random attack scenario as Fig.~\ref{fig:disturb_random}, the scale-free nature of the network renders both the baseline and the protected models virtually impressive, maintaining $\sim$99\% structural health. Because random attacks overwhelmingly strike peripheral leaf nodes (which have an out-degree of zero), the network's core remains untouched. This indicates that uniform, ecosystem-wide security protection is economically wasteful. What should be clarified is that the visible gap in absolute LCC size between Proposal A and Combined A+B is an artifact of Proposal B's pre-emptive edge pruning, rather than a result of attack degradation. Proposal B's edge pruning disconnects single-dependency leaf packages before the attack begins, lowering the starting LCC but eliminating structural fragility at those edges.

\begin{figure}[!t]
\centering
\includegraphics[width=\columnwidth]{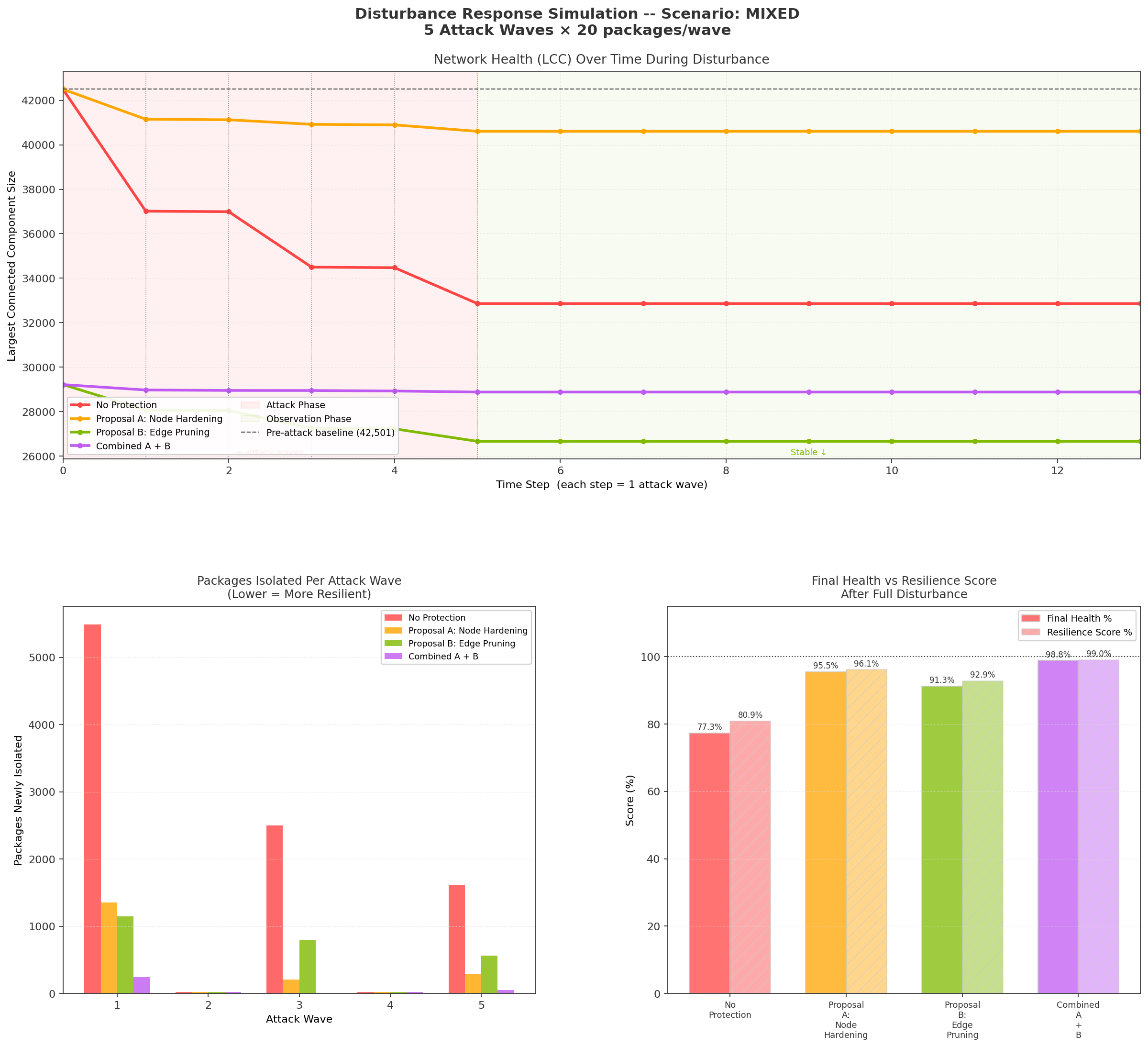}
\caption{Network response to a mixed attack. Waves 1, 3, and 5 target core hubs and waves 2 and 4 apply random removal, resulting in near-zero collateral isolation.}
\label{fig:disturb_mixed}
\end{figure}

\textbf{Adaptive Mixed Attacks.} To make the simulation more realistic and adaptive, we design a "Mixed" attack scenario that alternates between targeted hub removal (Waves 1, 3, 5) and uniformly random removal (Waves 2, 4), which can be seen in Fig.~\ref{fig:disturb_mixed}. The results perfectly illustrate the extreme topological inequality of the npm ecosystem. During the hub-targeted waves, the baseline network suffers catastrophic collateral isolation. However, during Waves 2 and 4, collateral damage drops to near-zero across all strategies. This proves that supply-chain fragility is almost entirely localized within the top 1\% of central packages; removing peripheral nodes yields virtually no cascading structural failure.

\begin{table}[!t]
\caption{Disturbance Response Results (5 Waves $\times$ 20 Packages)}
\label{tab:disturbance}
\centering
\begin{tabular}{llrrr}
\toprule
\textbf{Scenario} & \textbf{Strategy} & \textbf{Health} & \textbf{Wave~1 Loss} \\
\midrule
\multirow{4}{*}{Targeted}
  & No Protection      & 71.7\% & $\approx$5{,}400     \\
  & Proposal A         & 94.7\% & $\approx$1{,}300  \\
  & Proposal B         & 87.9\% & $\approx$1{,}100     \\
  & Combined           & \textbf{98.7\%} & $\approx$200     \\
\midrule
\multirow{2}{*}{Random}
  & No Protection      & 99.3\% & $\approx$20 \\
  & Proposal A         & 99.5\% & $\approx$20  \\
\midrule
\multirow{4}{*}{Mixed}
  & No Protection      & 77.3\% & $\approx$5{,}400      \\
  & Proposal A         & 95.5\% & $\approx$1{,}400  \\
  & Proposal B         & 91.3\% & $\approx$1{,}100  \\
  & Combined           & \textbf{98.8\%} & $\approx$200    \\
\bottomrule
\end{tabular}
\begin{tablenotes}
    \footnotesize
    \item[*] With $p_{breach}=0.15$, no individual wave is fully absorbed, so the column is omitted.
\end{tablenotes}
\end{table}

\subsection{Threat Vector Analysis}

\begin{figure}[!t]
\centering
\includegraphics[width=\columnwidth]{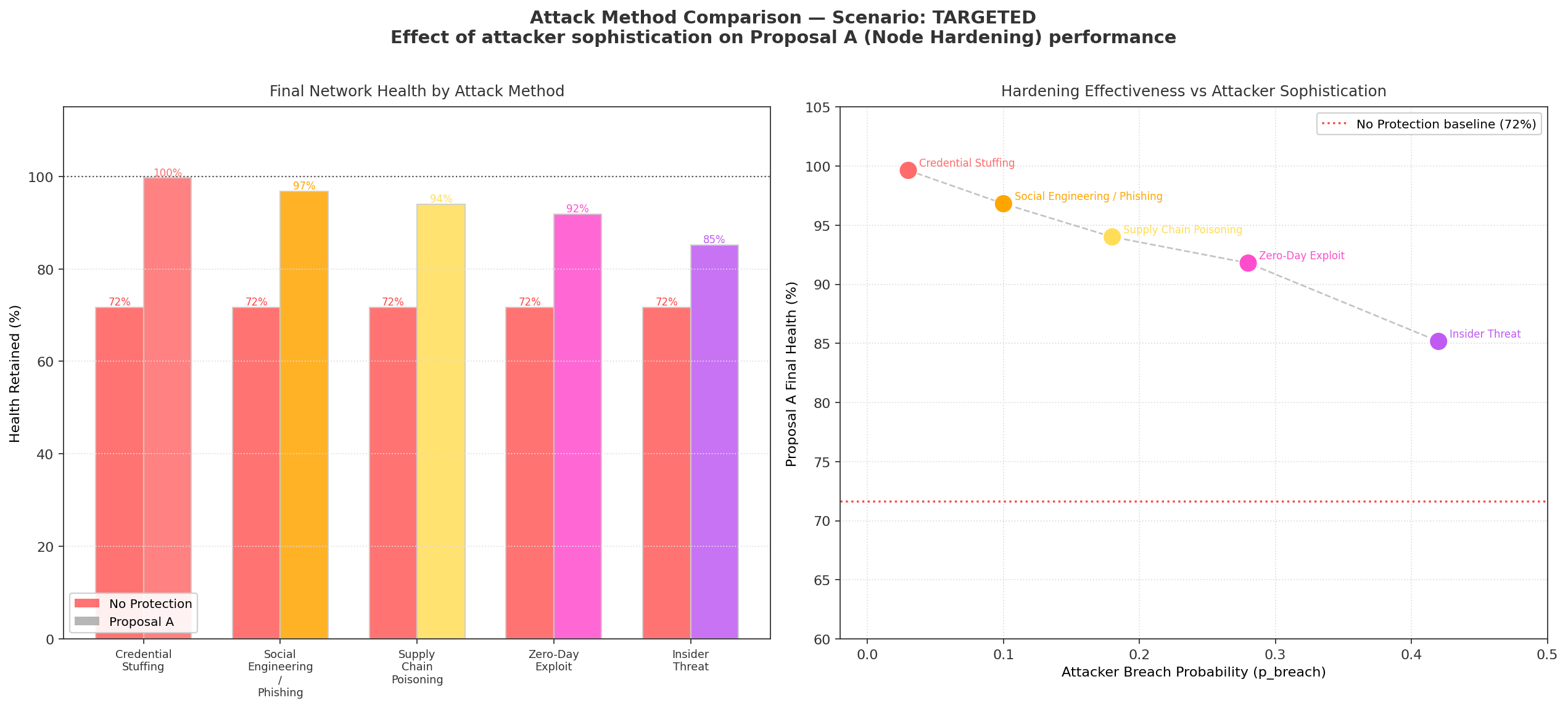}
\caption{Network response to escalating real-world threat vectors. The effectiveness of Centrality-Based Hardening scales inversely with attacker sophistication ($p_{breach}$).}
\label{fig:threat_vectors}
\end{figure}

\begin{figure}[!t]
\centering
\includegraphics[width=\columnwidth]{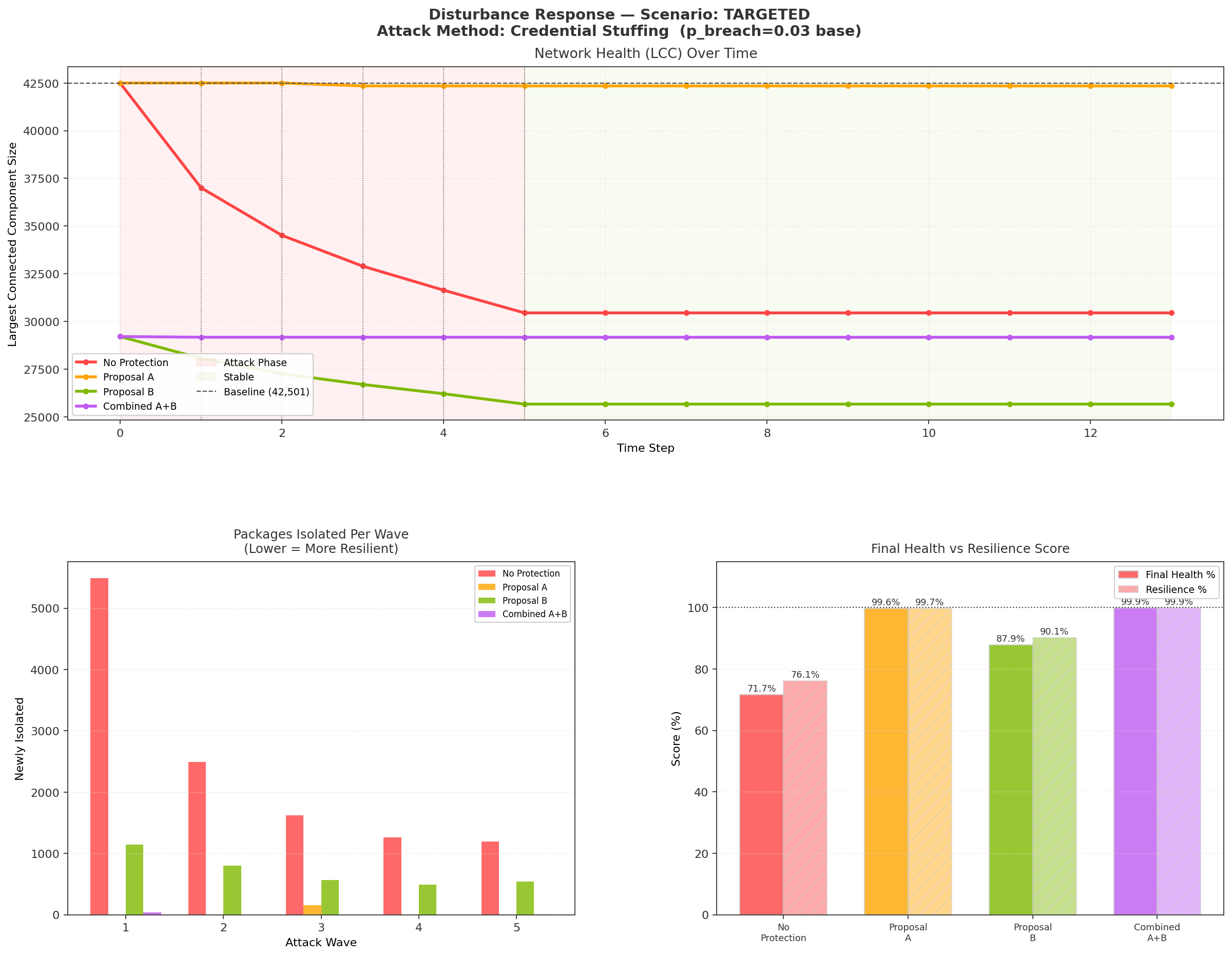}
\caption{Disturbance response isolated for the Credential Stuffing threat vector ($p_{breach} = 0.03$).}
\label{fig:supp_cred}
\end{figure}

To model these defenses realistically, we evaluate Proposal A against specific Threat Vectors with escalating breach probabilities ($p_{breach}$), as shown in Fig.~\ref{fig:threat_vectors}. Against automated threats such as "Credential Stuffing" ($p_{breach} = 0.03$), the mandatory 2FA protocols of Proposal A effectively create an impenetrable wall, maintaining near 100\% network health (shown in Fig.~\ref{fig:supp_cred}). 

\begin{figure}[!t]
\centering
\includegraphics[width=\columnwidth]{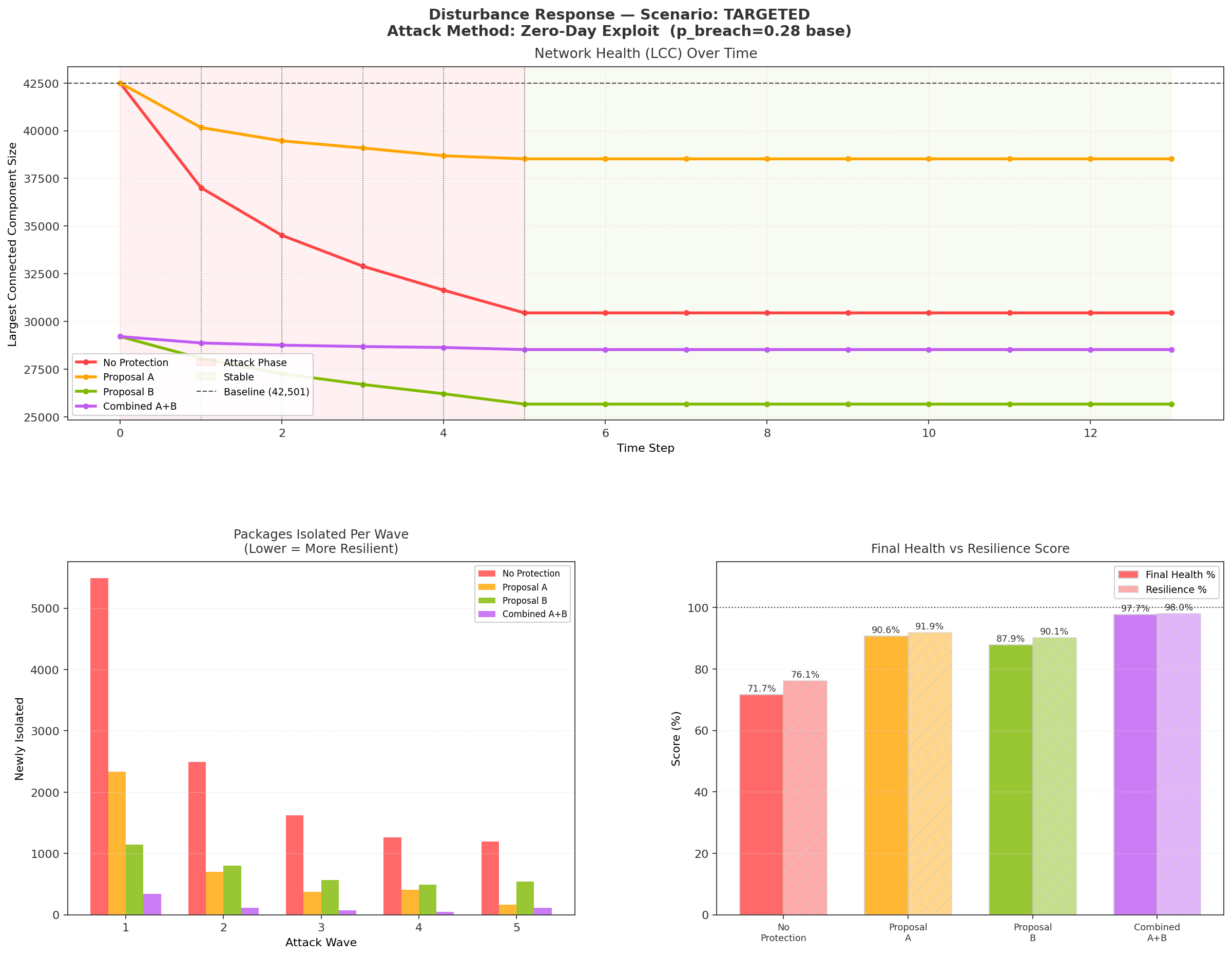}
\caption{Disturbance response isolated for the Zero-Day Exploit threat vector ($p_{breach} = 0.28$).}
\label{fig:supp_zero}
\end{figure}

However, against a sophisticated "Zero-Day Exploit"($p_{breach} = 0.28$) or "Insider Threat" ($p_{breach} = 0.42$), protection predictably degrades: Proposal A retains 90.6\% health against the former and 85\% against the latter, which can be seen in Fig.\ref{fig:supp_zero}. This demonstrates the realistic limits of centrality-based defense: while structural policies can neutralize automated supply-chain poisoning, mitigating authorized, malicious insiders requires deeper semantic code analysis beyond the scope of pure network topology.

\subsection{Evaluation Metrics}
\label{subsec:metrics}

To provide a rigorous multi-dimensional evaluation beyond scenario simulations, we define twelve metrics across two categories.

\begin{figure*}[!t]
\centering
\includegraphics[width=\textwidth]{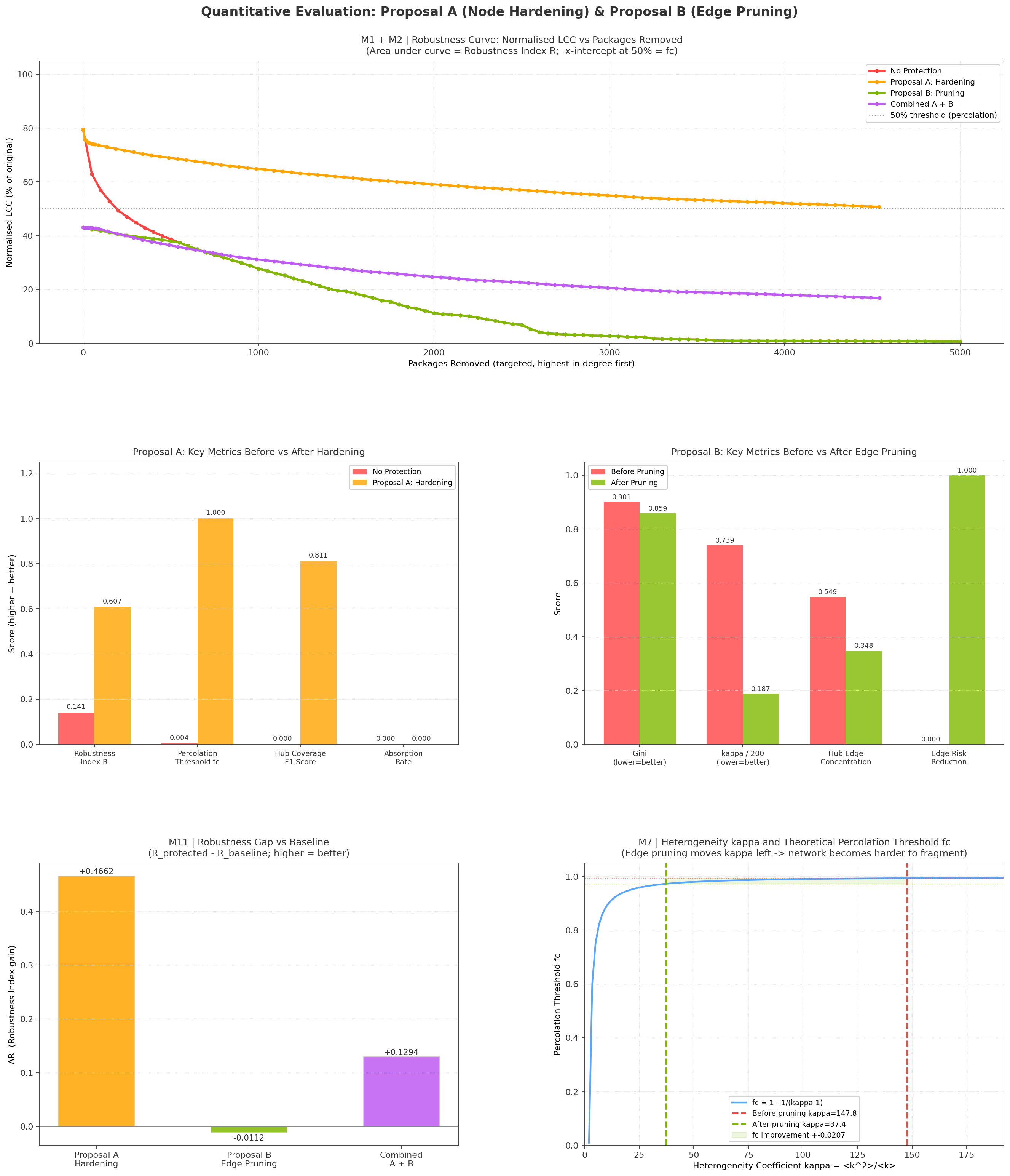}
\caption{Quantitative evaluation dashboard (M1--M11). \textit{Upper}: Robustness curves (M1+M2) for all four conditions over 5{,}000 targeted removals. \textit{Middle-left}: Proposal~A metrics before/after hardening. \textit{Middle-right}: Proposal~B metrics before/after pruning. \textit{Lower-left}: Robustness gap M11 for all three protected strategies. \textit{Lower-right}: Heterogeneity $\kappa$ and theoretical percolation threshold (M7)---edge pruning moves $\kappa$ from 147.8 to 37.4.}
\label{fig:metrics}
\end{figure*}

\textbf{Proposal A metrics (M1--M5).}
The \textit{robustness index}~\cite{schneider2011}:
\begin{equation}
  R = \frac{1}{N} \sum_{q} \frac{\mathcal{H}(G_q)}{N}
  \label{eq:R}
\end{equation}
measures the normalised area under the LCC curve as nodes are removed sequentially. The \textit{percolation threshold} $f_c$ is the fraction of nodes removed when $\mathcal{H}(G)/N$ first drops below 0.5. \textit{Hub coverage quality} is measured by the F1 score of the in-degree-based Critical Node selection against a PageRank-based ground truth. \textit{Attack absorption rate} is the fraction of attack waves in which zero packages are isolated. \textit{Mean LCC drop per wave} is the average packages newly isolated per wave.

\textbf{Proposal B metrics (M6--M10).}
Beyond the Gini and $\kappa$ metrics already introduced (Eqs.~\ref{eq:gini},~\ref{eq:kappa}), we measure \textit{natural connectivity}~\cite{jun2010}:
\begin{equation}
  \bar{\lambda} = \ln \left( \frac{1}{N} \sum_{i=1}^{N} e^{\lambda_i} \right)
  \label{eq:nat_conn}
\end{equation}
where $\lambda_i$ are the eigenvalues of the adjacency matrix of the top-$k$ hub subgraph (computed on a 800-node sample for tractability). The \textit{hub edge concentration ratio} $\eta$ (Eq.~\ref{eq:hub_conc}) and \textit{edge risk reduction rate} complete the structural metrics.

\begin{figure}[!t]
\centering
\includegraphics[width=\columnwidth]{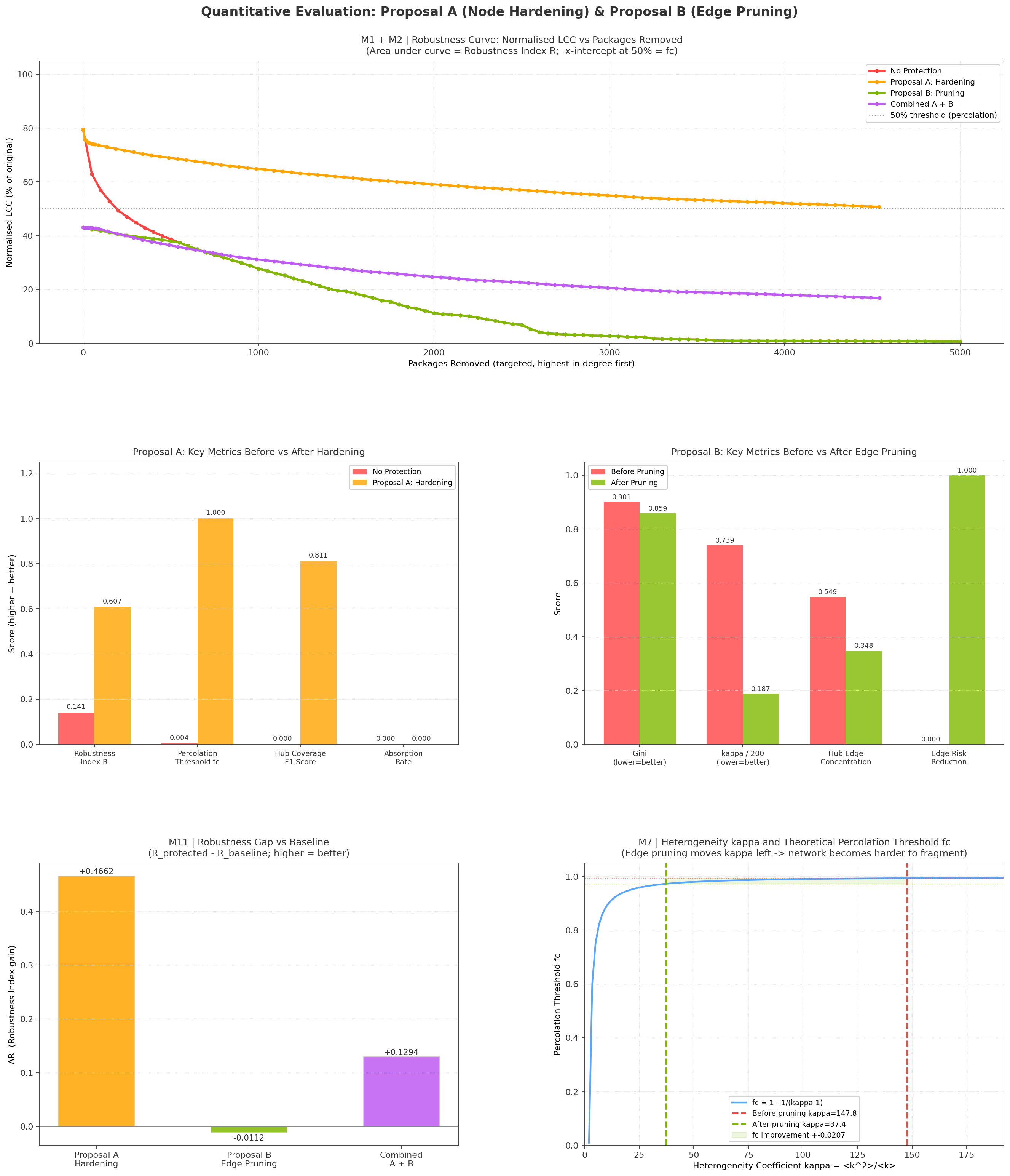}
\caption{Overall evaluation metrics comparing the structural Robustness Index ($R$) and Cascade Reduction efficiencies across all proposed strategies.}
\label{fig:eval_metrics}
\end{figure}

Table~\ref{tab:metrics} presents all results. Fig.~\ref{fig:metrics} visualises the five most informative metrics.

\begin{table}[!t]
\caption{Quantitative Evaluation Metrics M1--M11}
\label{tab:metrics}
\centering
\begin{tabular}{clrrr}
\toprule
\textbf{ID} & \textbf{Metric} &
\textbf{Before} & \textbf{After} & \textbf{Change}\\
\midrule
M1  & Robustness index $R$        & 0.141  & 0.607 & $+0.466$\\
M2  & Percolation threshold $f_c$ & 0.4\%  & 100\% & $+99.6$pp\\
M3  & Hub coverage F1             & ---    & 0.811 & ---\\
M4  & Absorption rate$^\dagger$   & 0\%    & ---   & ---\\
M5  & LCC drop / wave             & 5{,}400 & $\sim$1{,}300 & $-76\%$\\
\midrule
M6  & Gini coefficient            & 0.901 & 0.859 & $-0.042$\\
M7  & $\kappa$                    & 147.8 & 37.4  & $-74.7\%$\\
M8  & Natural connectivity        & 0.131 & 0.089 & $-0.042$\\
M9  & Hub edge conc.\ $\eta$      & 54.9\% & 34.8\% & $-36.6\%$\\
M10 & Edge risk reduction         & 24{,}149 & 0 remain & 100\%\\
\midrule
M11 & Rob.\ gap (A)   & \multicolumn{2}{c}{$R_A-R_0$}  & $+0.466$\\
M11 & Rob.\ gap (B)   & \multicolumn{2}{c}{$R_B-R_0$}  & $-0.011^*$\\
M11 & Rob.\ gap (A+B) & \multicolumn{2}{c}{$R_{AB}-R_0$} & $+0.129$\\
\bottomrule
\multicolumn{5}{l}{\small $^\dagger$Under ideal hardening
  ($p_{breach}=0$): 100\%; under realistic}\\
\multicolumn{5}{l}{\small $\quad p_{breach}=0.15$: $\approx0\%$ (some nodes breached each wave).}\\
\multicolumn{5}{l}{\small $^*$Negative due to leaf-node
  disconnection; see \S\ref{sec:concl}.}
\end{tabular}
\end{table}

Ultimately, the quantitative success of our dual-pronged approach exhibits the effectiveness of our combined strategy. Taken together, the metrics confirm that Proposal A raises $R$ by +0.466 and eliminates the percolation risk ($f_c: 0.4\% \to 100\% $), while Proposal B reduces structural load concentration by 36.6\% and $\kappa$ by 74.7\%, providing complementary long-term resilience. By combining Centrality-Based Node Hardening with Dependency Weight Warning (Edge Pruning), the network achieves a paradigm where structural robustness is preserved while malware contagion pathways are systematically neutralized.

\section{Conclusion and Limitation}
\label{sec:concl}
We study the npm ecosystem and demonstrate that the npm dependency graph is a scale-free network, whose topology makes it catastrophically fragile to targeted hub attacks while remaining robust to random failures. To handle this vulnerability, we propose and evaluate the Targeted Node-Hardening Protocol (TNHP), a dual-mechanism defence grounded directly in the network's structural properties. This mechanism includes the Centrality-Based Node Hardening and Dependency Weight Warning system. Through rigorous simulation of multi-wave attack scenarios and a comprehensive set of quantitative metrics, we show that TNHP can significantly reduce malware cascade size while maintaining near-perfect network health under targeted attacks. Our work provides a scalable, data-driven strategy for securing software supply chains by strategically strengthening the most critical nodes.
However, there are limitations to our approach. First, while our composite metric successfully incorporates both in-degree and PageRank, it may still overlook specific topological bridges, namely those nodes with moderate degrees but extremely high out-degree that connect disparate ecosystems. Second, while our simulations model various attack scenarios, real-world attacks may involve more complex strategies that adapt to defenses, such as multi-vector attacks or insider threats. Finally, the implementation of mandatory security measures for critical nodes may face practical challenges in terms of developer compliance and ecosystem governance, which warrants further investigation into incentive structures and community engagement strategies.

\vfill

\end{document}